\documentclass{article}

\usepackage{amssymb,amsfonts,amsmath}
\usepackage{cite,enumerate,float,indentfirst}
\usepackage{color}

\def\be{\begin{eqnarray}}
\def\ee{\end{eqnarray}}
\def\nn{\nonumber}

\def\p{\partial}
\def\tr{{\rm tr}\,}
\def\Tr{{\rm Tr}\,}

\definecolor{red}{rgb}{1,0,0}
\definecolor{orange}{rgb}{1,0.5,0}
\definecolor{violet}{rgb}{0.7,0,1}

\hoffset= - 1.0in         

\begin{document}

\title{\vspace{1.1cm}{\LARGE {\bf Complete solution to Gaussian tensor model\\
and its integrable properties
}\vspace{.5cm}}
\author{{\bf H. Itoyama$^{a,b,c}$},
{\bf A. Mironov$^{d,e,f}$},
\ {\bf A. Morozov$^{g,e,f}$}
}
\date{ }
}

\maketitle

\vspace{-7.2cm}

\begin{center}
\hfill FIAN/TD-11/19\\
\hfill ITEP/TH-21/19\\
\hfill IITP/TH-13/19\\
\hfill MIPT/TH-11/19\\
\hfill OCU-PHYS-511\\
\hfill NITEP 34
\end{center}

\vspace{4.cm}

\noindent
$^a$ {\small {\it Nambu Yoichiro Institute of Theoretical and Experimental Physics (NITEP) and}} \\
\phantom{a} {\small {\it Department of Mathematics and Physics, Graduate School of Science,}}  \\
\phantom{a} {\small {\it Osaka City University, 3-3-138, Sugimoto, Sumiyoshi-ku, Osaka, 558-8585, Japan  }}\\
$^b$ {\small {\it I.E.Tamm Theory Department, Lebedev Physics Institute, Leninsky prospect, 53, Moscow 119991, Russia}}\\
$^c$ {\small {\it ITEP, B. Cheremushkinskaya, 25, Moscow, 117259, Russia }}\hfill\\
$^d$ {\small {\it Institute for Information Transmission Problems,  Bolshoy Karetny per. 19, build.1, Moscow 127051 Russia}}\\
$^e$ {\small {\it MIPT, Dolgoprudny, 141701, Russia}}\\

\vspace{.5cm}

\begin{abstract}
 Similarly to the complex matrix model, the rainbow tensor models are superintegrable in the sense that arbitrary Gaussian correlators are explicitly expressed through the Clebsh-Gordan coefficients.  We introduce associated (Ooguri-Vafa type) partition functions and describe their $W$-representations. We also discuss their integrability properties, which can be further improved by better adjusting the way the partition function is defined. This is a new avatar of  the old unresolved problem with non-Abelian integrability concerning a clever choice of the partition function. This is a part of the long-standing problem to define a non-Abelian lift of integrability from the fundamental to generic representation families of arbitrary Lie algebras.
\end{abstract}

\bigskip

\bigskip

\paragraph{Introduction.} For Gaussian measures, it is possible to find
an {\it explicit full} basis of gauge invariant observables,
which have {\it factorized} averages
and can be written in the form of {\it explicit rational}
functions of matrix size \cite{MM}.
This means that matrix models are not just integrable,
i.e. expressed through distinguished, still transcendental
$\tau$-functions \cite{GMMMO,KMMOZ,versus},
but {\bf  super-integrable} like particles moving
in especially nice potentials, say, in oscillator or Coulomb ones.
This special basis is actually formed by ``characters",
which are the Schur or Macdonald polynomials \cite{Mac}
(in the case of $q,t$-deformed models \cite{MPSh}).
A similar property persists for logarithmic (hypergeometric)
measures \cite{MMAGT} when Selberg integrals convert the generalized Macdonald
polynomials \cite{genM} into factorized Nekrasov functions \cite{Nek},
this fact is used in the conformal matrix model \cite{CMM}
proof \cite{MMSh} of the AGT relations \cite{AGT}.

In this letter, we explain what happens in still another
generalization: from matrices to tensors \cite{tenmods,tenmodsSYK,Ramg1,Ramg2,IMMgcb}.
As explained in \cite{IMMgc}, in this case of Gaussian measure,
there is a large {\it kernel}, still the rainbow tensor models \cite{IMMrain}
remain {\bf super-integrable}  in the sense that
beyond the kernel one can still find an explicit basis
with nicely factorized and explicitly calculable averages.

This poses a further question of what super-integrability
in the above sense implies for the ordinary integrability.
The latter is usually seen at two levels:
as an infinite set of linear Ward identities
(Virasoro-like constraints) \cite{Vir}
and as a set of bilinear Hirota-like equations \cite{MMintrev}.
In the both cases, one needs an additional input,
an appropriately defined generating function \cite{GKLMM}
of averages, which is then identified as a $\tau$-function
subject to an additional string/Painleve constraint
(this peculiar class is called ``matrix-model $\tau$-functions) \cite{FKN,GKMMO,KS,GKM,MMZ2}.
A natural choice of partition function for the correlators of characters
are Cauchy sums with weights which are also characters,
in physical literature they are also known as Ooguri-Vafa
partition functions (since they were used in the widely known paper \cite{OV}).
We demonstrate that, in the tensor case, these Ooguri-Vafa
partition functions have rich,
still limited integrability properties,
which can stimulate a new attention to \cite{GKLMM} and a search
for a somewhat better prescription for making the generating functions.

\paragraph{Gaussian tensor models.}

The Gaussian tensor model is a model of complex $r$-tensors $M_{a^1,...a^r}$ with the Gaussian action
\be
S:=\Tr M_{a^1,...a^r}\bar M^{a^1,...a^r}
\ee
i.e. the averages in this model are given by
\be
\Big<\ldots \Big> = {\int e^{-\Tr M\bar M}\ldots d^2M\over \int e^{-\Tr M\bar M} d^2M},
\ \ \ \ \ \ \ \ {\rm with} \ \ \ \ \ \
d^2M = \prod_{i=1}^r\prod_{a^i=1}^{N_i} d^2M_{a^1,...a^r}
\ee
The gauge invariant operators at the level $n$ in any (not obligatory Gaussian) tensor model are the tensorial counterparts of ``multi-trace" operators
\be\label{gio}
{\cal K}^{(n)}_{\sigma_1,\ldots,\sigma_r}
= \sum_{\vec a^1=1}^{N_1}\ldots\sum_{\vec a^r=1}^{N_r}
\left(\prod_{p=1}^n M_{a^1_p,...a^r_p}
\bar M^{a^1_{\sigma_1(p)},\ldots,a^r_{\sigma_r(p)}}\right)
\ee
where $\sigma_m$ are elements of the permutation group $S_n$. Note that the operators ${\cal K}$ are invariant w.r.t. simultaneous multiplying all $\sigma_i$ by an arbitrary element $\gamma$ of the symmetric group: $\sigma_i\to\sigma_i\circ\gamma$. This allows one to consider only ${\cal K}^{(n)}_{id,\sigma_2,\ldots,\sigma_r}$ without any loss of generality.
Even after this, there is still an invariance w.r.t. simultaneous conjugation of all $\sigma_i$ by an arbitrary element $\zeta$ of the symmetric group: $\sigma_i\to\zeta\circ\sigma_i\circ\zeta^{-1}$.

There is also a distinguished set of operators, which were called generalized characters in \cite{IMMgc} that are defined as
\be
\chi_{R_1,\ldots, R_r}(M,\bar M)
= \frac{1}{n!}\!\!\!\!\!\!\!\!\!\sum_{\ \ \ \ \ \sigma_1,\ldots,\sigma_r\in S_n}\!\!\!\!\!\!\!\!\!
\psi_{R_1}(\sigma_1)\ldots\psi_{R_r}(\sigma_r)\cdot {\cal K}^{(n)}_{\sigma_1,\ldots,\sigma_r}
\label{hurchar}
\ee
where $R_i$ are the Young diagrams (partitions), and $\psi_R(\sigma)$ is a character of symmetric group $S_n$ that actually depends only on the conjugacy class of $\sigma$. These generalized characters do not form a full basis in the space of all gauge invariant operators, but they form a over-complete basis in the space of all gauge invariant operators with non-vanishing Gaussian averages.

In the case of $r=2$, i.e. in the case of matrix model, the generalized characters reduce to the ordinary characters of the linear group \cite{IMMgc},
\be\label{gc2}
\chi_{R_1,R_2} \equiv   \frac{1}{n!}\sum_{\sigma_1,\sigma_2\in S_n}
\psi_{R_1}(\sigma_1)\psi_{R_2}(\sigma_2){\cal K}_{\sigma_1,\sigma_2}
= \frac{\delta_{R_1,R_2}}{d_{R_1}} \chi_{R_2}
\ee
hence, the name. Here $\chi_R\{p_k\}$ is the Schur function (the character of linear group) as a function of time-variables $p_k$. These time-variables are sometimes realized as the traces in the matrix model case, $p_k=\Tr\Big(M\bar M\Big)^k$, the monomials of traces being all gauge invariant operators (\ref{gio}) in this case.

\paragraph{Partition function.}

Since the generalized characters form a basis in the space of all gauge invariant operators with non-vanishing Gaussian averages, it is natural to choose as the generating function of all correlators the sum
\be\label{Z}\boxed{
Z_r\{p^{(i)}\}: = \sum_{R_1,\ldots,R_r} \prod_j\chi_{R_j}\{p^{(j)}\}\cdot\left<\chi_{_{R_1,\ldots,R_r}}  \right>}
\ee
In the simplest case of the matrix model $r=2$, one would have to consider
\be
Z_2\{p^{(1)},p^{(2)}\}: = \sum_{R_1,\ldots,R_r} \chi_{R_1}\{p^{(1)}\}\chi_{R_2}\{p^{(2)}\}
\cdot\left<\chi_{_{R_1,R_2}}  \right>
\ee
However, as we emphasize above, the basis of the generalized characters is over-complete. In this simplest case, the redundant contributions can be easily removed by considering $p^{(2)}=\delta_{1,k}$, with $d_R:=\chi_R\{p_k=\delta_{1,k}\}$:
\be\label{mamo}
Z_2\{p^{(1)}_k\}=\sum_{R,R_2} d_{R_2}
\chi_R\{p^{(1)}\}\cdot\left<\chi_{_{R_1,R_2}}  \right>\stackrel{(\ref{gc2})}{=}\sum_n{1\over n!}\sum_{R,R_2\vdash n} d_{R_2}
\chi_R\{p^{(1)}\}\sum_{\sigma_1,\sigma_2\in S_n}\psi_R(\sigma_1)\psi_{R_2}(\sigma_2)\left<{\cal K}_{\sigma_1,\sigma_2}\right>
=\nn\\
=\sum_n{1\over n!}\sum_{\sigma_1\in S_n}p^{(1)}_{[\sigma_1]}\left<{\cal K}_{\sigma_1,id}\right>
=\sum_n\sum_{\Delta\vdash n}{p^{(1)}_{\Delta[\sigma]}\over z_{\Delta[\sigma]}}\left<{\cal K}_{\sigma,id}\right>=
\left<\sum_\Delta{p^{(1)}_\Delta K_\Delta\over z_\Delta}\right>=\left<\exp\left(\sum_k{p^{(1)}_k\Tr(M\bar M)^k\over k}\right)\right>
\ee
where $[\sigma]$ denotes the cycle type of the permutation $\sigma$ and $\Delta[\sigma]$ the corresponding conjugacy class (which consists of ${n!\over z_\Delta}$ elements). We use the following notation: for the Young diagram $\Delta=\{\delta_1\geq \delta_2\geq\ldots\delta_{l_\Delta}>0\}
= \{1^{m_1},2^{m_2},\ldots\}$,
the symmetry factor is defined $z_\Delta:=\prod_i m_i! \cdot i^{m_i}$,
and $p_\Delta$ is  a monomial
$p_\Delta \equiv p_{\delta_1}p_{\delta_2}\ldots p_{\delta_{l }}$.
We also used that ${\cal K}_{\sigma,id}$ depends only on the conjugacy class of $\sigma$ and the formulas \cite{SF}
\be
\sum_\Delta{p_\Delta
p'_\Delta\over z_\Delta}=\exp\left(\sum_k{p_kp'_k\over k}\right)\\
\sum_R\chi_R\{p_k\}\psi_R(\sigma)=p_{[\sigma]}
\ee
Formula (\ref{mamo}) is the standard generating function of correlators in the matrix model case,

In the generic tensor model, there is no simple way to remove the redundancy. For instance, in the $r=3$ case, consider
\be\label{nr3}
Z_3\{p^{(1)}_k,p^{(2)}_k\}:= \sum_{R_1,R_2,R_3}
\chi_{R_1}\{p^{(1)}\}\chi_{R_2}\{p^{(2)}\}d_{R_3}\cdot\left<\chi_{_{R_1,R_2,R_3}}  \right>\stackrel{(\ref{hurchar})}{=}\nn\\
=\sum_n{1\over n!}\sum_{R_1,R_2,R_3\vdash n} \chi_{R_1}\{p^{(1)}\}\chi_{R_2}\{p^{(2)}\}d_{R_3}
\sum_{\sigma_1,\sigma_2,\sigma_3\in S_n}\psi_{R_1}(\sigma_1)\psi_{R_2}(\sigma_2)\psi_{R_3}(\sigma_3)
\left<{\cal K}_{\sigma_1,\sigma_2,\sigma_3}\right>
=\nn\\
=\sum_n{1\over n!}\sum_{\sigma_1,\sigma_2\in S_n}p^{(1)}_{[\sigma_1]}p^{(2)}_{[\sigma_2]}\left<{\cal K}_{\sigma_1,\sigma_2,id}\right>
\ee
Since the time variables depends only on two conjugacy classes, $[\sigma_1]$ and $[\sigma_2]$, and the operator ${\cal K}_{\sigma_1,\sigma_2,id}$ generically depends not only on these two classes, this partition function is not enough, and one has to consider the complete partition function $Z_3\{p^{(i)}\}$, (\ref{Z}), which depends on three sets of time variables.

\paragraph{Complete solution to the Gaussian tensor model.}

The averages of these generalized characters are equal to \cite{MMten}
\be\label{Ga}
\Big< \chi_{R_1,\ldots, R_r} \Big> \ =
\ C_{R_1,\ldots,R_r} \cdot
\frac{D_{R_1}(N_1)\cdot \ldots\cdot D_{R_r}(N_r)}{d_{R_1}\cdot\ldots \cdot d_{R_r}}
\ee
where $D_R(N)$ is the dimension of the linear group, $D_R(N)=\chi_R\{p_k=N\}$, and
\be
C_{R_1,\ldots,R_r}:=\sum_{\Delta\vdash n}{\prod_{i=1}^r \psi_{R_i}(\Delta)\over z_\Delta}
\ee
In the case of $r=3$, $C_{R_1,R_2,R_3}$ are the Clebsch-Gordan coefficients of the three irreducible representations $R_1$, $R_2$, $R_3$ of the symmetric group.

Note that
\be\label{DdR}
\frac{D_R(N)}{d_R} = \prod_{(i,j)\in R} (N+i-j)
\ee
and one can finally write down the complete solution to the Gaussian tensor model, that is, the explicit formula for the partition function (\ref{Z}):
\be\label{Zr}\boxed{
\begin{array}{c}
\displaystyle{Z_r\{p^{(i)}\}: = \sum_{R_1,\ldots,R_r} \prod_j\chi_{R_j}\{p^{(j)}\}\cdot
\ C_{R_1,\ldots,R_r} \cdot
\frac{D_{R_1}(N_1)\cdot \ldots\cdot D_{R_r}(N_r)}{d_{R_1}\cdot\ldots \cdot d_{R_r}}
=}\\
\\
\displaystyle{=\sum_{R_1,\ldots,R_r}C_{_{R_1\ldots R_r}} \prod_{m=1}^r\left(\chi_{R_m}\{p^{(m)}\}
\cdot\prod_{(i,j)\in R_m} (N_m+i-j)\right)}
\end{array}}
\ee
Note that, when the last product in (\ref{Zr}) is constant (for instance, at all $N_m$ large), this function is equal to \cite{IMMgc}
\be\label{ttau}
Z_r\{p^{(i)}\}\sim\sum_{R_1,\ldots,R_r}C_{_{R_1\ldots R_r}} \prod_{m=1}^r\chi_{R_m}\{p^{(m)}\}
=\exp\left(\sum_k
{\prod_{m=1}^rp^{(m)}_k\over k}\right)
\ee
Here we used a generalization of the Cauchy formula to multi-linear sums of characters \cite{IMMgc}.
It gives a trivial $\tau$-function of the KP hierarchy w.r.t. to any of the sets of times $p_k/k$.

\paragraph{Integrability.}

In order for a linear combination of the Schur functions,
\be\label{tP}
\tau\{p_k\}=\sum_Rw_R\chi_R\{p_k\}
\ee
to be a $\tau$-function of the KP hierarchy, the expansion coefficients $w_R$ have to satisfy the so called Pl\"ucker relations \cite{DJKM,Sat}, which are best written in terms of
\textit{the Frobenius variables} describing the Young diagram $R$ (the first row is the horizontal leg lengths of hooks,
and the second row is the vertical leg lengths of the corresponding hooks):
\be
w\left(\begin{array}{c}
i_1\ldots\check i_\mu\ldots\check i_\nu\ldots i_r\\
j_1\ldots\check j_\mu\ldots\check j_\nu\ldots j_r
\end{array}\right)
w\left(\begin{array}{c} i_1\ldots i_r \\ j_1\ldots j_r\end{array}\right)
- w\left(\begin{array}{c}
i_1\ldots\check i_\mu\ldots i_r\\
j_1\ldots\check j_\mu\ldots j_r
\end{array}\right)
w\left(\begin{array}{c}
i_1\ldots\check i_\nu\ldots i_r\\
j_1\ldots\check j_\nu\ldots j_r
\end{array}\right) + \nn \\
+ w\left(\begin{array}{c}
i_1\ldots\check i_\mu\ldots i_r\\
j_1\ldots\check j_\nu\ldots j_r
\end{array}\right)
w\left(\begin{array}{c}
i_1\ldots\check i_\nu\ldots i_r\\
j_1\ldots\check j_\mu\ldots j_r
\end{array}\right) = 0
\ee
The first few relations in the explicit form are
\be
\begin{array}{|c|}
\hline
w_{22}w_0 - w_{21}w_1 + w_2w_{11} = 0, \nn \\
\hline
w_{32}w_0 - w_{31}w_1 + w_3w_{11} = 0, \nn \\
w_{221}w_0 - w_{211}w_1 + w_2w_{111} = 0, \nn \\
\hline
w_{42}w_0 - w_{41}w_1 + w_4w_{11} = 0, \nn \\
w_{33}w_0 - w_{31}w_2 + w_3w_{21} = 0, \nn \\
w_{321}w_0 - w_{311}w_1 + w_3w_{111} = 0, \nn \\
w_{222}w_0 - w_{211}w_{11} + w_{21}w_{111} = 0, \nn \\
w_{2211}w_0 - w_{2111}w_1 + w_2w_{1111} = 0, \nn \\
\hline
\ldots
\end{array}
\ee
Now note that any exponential linear in times, $\exp\Big(\sum_ka_kp_k\Big)$ is a trivial $\tau$-function of the KP hierarchy, because it can be expanded using the Cauchy formula into the bilinear combination (\ref{tP}) with $w_R=\chi_R\{ka_k\}$, which satisfy the Pl\"ucker relations. Thus, from the generalized Cauchy formula (\ref{ttau}) it follows that
\be
\tau^{(r)}\{p^{(1)}_k,\ldots,p^{(r)}_k\}:=\sum_{R_1,\ldots,R_r}C_{_{R_1\ldots R_r}} \prod_{m=1}^r\chi_{R_m}\{p^{(m)}\}
\ee
is a (trivial) KP $\tau$-function with respect to any of times $p_k/k$. For the definiteness, we distinguish the first set of times, $p^{(1)}_k$ and consider further integrability with respect to this set of times. Then, the Pl\"ucker coordinates are
\be\label{ttau2}
w_R=\sum_{R_2,\ldots,R_r}C_{_{R,R_2\ldots R_r}} \prod_{m=2}^r\chi_{R_m}\{p^{(m)}\}
\ee
Now note that any function $w_R$ that satisfies the Pl\"ucker relations can be multiplied $w_R\longrightarrow w_R\prod_{i,j\in R}f(i-j)$ by an arbitrary function $f(x)$ and still continues to satisfy the Pl\"ucker relations \cite{gtau}: these latter are invariant with respect to this operation. This means that one can multiply (\ref{ttau2}) by
\be
\frac{D_R(N)}{d_R}
\ee
and, due to (\ref{DdR}), it still will be a $\tau$-function:
\be\label{tau2}
\tau_0\{p^{(1)}_k,\ldots,p^{(r)}_k\}=\sum_{R_1,\ldots,R_r}\frac{D_{R_1}(N)}{d_{R_1}}C_{_{R_1\ldots R_r}} \prod_{m=1}^r\chi_{R_m}\{p^{(m)}\}
\ee
Unfortunately, it will be no longer a $\tau$-function after further multiplying the summand in (\ref{ttau2}) by
\be
\prod_{i=2}^r\frac{D_R(N)}{d_R}
\ee
and, hence, (\ref{Z}),
\be\label{rtau}
Z_r\{p^{(i)}\}: = \sum_{R_1,\ldots,R_r} \prod_j\chi_{R_j}\{p^{(j)}\}\cdot\left<\chi_{_{R_1,\ldots,R_r}}  \right>
\ee
is not a $\tau$-function of the KP hierarchy. It becomes such only if one restricts all the sets of time variables with $i=2,\ldots, r$: $p^{(i)}_k=\delta_{1,k}$, since then (\ref{rtau}) is just
\be
\tau^{(r)}\{p^{(1)}_k\}=\sum_{R,R_2,\ldots,R_r} \left(\prod_{j=2}^r d_{R_j}\right)
\chi_R\{p^{(1)}\}\cdot\left<\chi_{_{R_1,\ldots,R_r}}  \right>=\sum_{R,R_2,\ldots,R_r} C_{R,R_2,\ldots,R_r}
\chi_R\{p^{(1)}\}{D_R(N_1)\over d_R}\prod_{i=2}^rD_{R_i}(N_i)
\ee
where we used that $d_R:=\chi_{_R}\{p_k=\delta_{k,1}\}$. Indeed, since $D_R(N)=\chi_R\{p_k=N\}$, it is just $\tau_0$, (\ref{tau2}) at all $p_k=N$.

Thus, we obtain that
\be\label{ti}
\tau^{(2)}\{p^{(1)}_k\}:= \sum_{R,R_2} d_{R_2}
\chi_R\{p^{(1)}\}\cdot\left<\chi_{_{R_1,R_2}}  \right>\stackrel{(\ref{Ga})}{=}\sum_R\chi_R\{p^{(1)}\}
{D_R(N_1)D_R(N_2)\over d_R}\nn\\
\tau^{(3)}\{p^{(1)}_k\}:= \sum_{R,R_2,R_3} d_{R_2}d_{R_3}
\chi_R\{p^{(1)}\}\cdot\left<\chi_{_{R,R_2,R_3}}  \right>\stackrel{(\ref{Ga})}{=}\sum_R\chi_R\{p^{(1)}\}C_{RR_2R_3}
{D_{R}(N_1)D_{R_2}(N_2)D_{R_3}(N_3)\over d_{R}}\nn\\
\ldots
\ee
are KP $\tau$-functions w.r.t. the time variables $p^{(1)}_k/k$,

In the case of $r=2$, it is sufficient to consider only this restricted set in order to generate all operators in the model, as we explained in (\ref{mamo}).

Unfortunately, in the tensor case, one can generate this way only a restricted set of necessary operators (those having non-vanishing Gaussian average). For instance, in the case of $r=3$,
\be
\tau^{(3)}\{p^{(1)}_k\}= \sum_{R,R_2,R_3} d_{R_2}d_{R_3}
\chi_R\{p^{(1)}\}\cdot\left<\chi_{_{R,R_2,R_3}}  \right>\stackrel{(\ref{hurchar})}{=}\nn\\
=\sum_n{1\over n!}\sum_{R,R_2\vdash n} d_{R_2}d_{R_3}
\chi_R\{p^{(1)}\}\sum_{\sigma_1,\sigma_2,\sigma_3\in S_n}\psi_R(\sigma_1)\psi_{R_2}(\sigma_2)\psi_{R_3}(\sigma_3)\left<{\cal K}_{\sigma_1,\sigma_2,\sigma_3}\right>
=\nn\\
=\sum_n{1\over n!}\sum_{\sigma_1\in S_n}p^{(1)}_{[\sigma_1]}\left<{\cal K}_{\sigma_1,id,id}\right>
=\left<\exp\left(\sum_k {p^{(1)}_k\tr (M\bar M)^k\over k}\right)\right>
\ee
where the product and the trace are understood here in the ordinary matrix sense for the matrices $M$, $\bar M$ with the first index being the usual matrix index, while the second and the third indices being combined into a single multi-index. This also follows from the
formula\footnote{This formula follows from the generalized Cauchy formula (\ref{ttau}) upon choosing $r=3$, $p^{(2)}_k=N_1$, $p^{(3)}_k=N_2$ and further expanding the r.h.s. of (\ref{ttau}) into the sum over $\chi_R\{p^{(1)}_k\}$ using the ordinary Cauchy formula.}
\be
\sum_{R_1,R_2}C_{R,R_1,R_2}D_{R_1}(N_1)D_{R_2}(N_2)=D_R(N_1N_2)
\ee
and (\ref{ti}).

The generalization to other $r$ is evident, with just last $r-1$ indices united into a multi-index. Unfortunately, the class of operators ${\cal  K}_{\sigma,id,\dots,id}$ is too small to generate all non-vanishing Gaussian averages in the tensor model with $r>2$.

Note also that, at $r=2$, one can form a partition function that depends on two sets of times, which is a $\tau$-function of the KP hierarchy
\be
\tau^{(2)}\{p^{(1)}_k,p^{(2)}_k\}:=Z_2\{p^{(1)}_k,p^{(2)}_k\}= \sum_{R_1,R_2}
\chi_{R_1}\{p^{(1)}\}\chi_R\{p^{(2)}\}\cdot\left<\chi_{_{R_1,R_2}}  \right>\stackrel{(\ref{Ga})}{=}
\sum_R\chi_R\{p^{(1)}\}\chi_R\{p^{(2)}\}{D_R(N_1)D_R(N_2)\over d_R^2}
\ee
In fact, it is a $\tau$-function of the Toda chain hierarchy.

Unfortunately, $Z_r\{p^{(1)}_k,p^{(2)}_k\}$ is not a KP $\tau$-function already in the case of $r=3$,
\be
Z_3\{p^{(1)}_k,p^{(2)}_k,\delta_{1,k}\}:= \sum_{R_1,R_2,R_3}
\chi_{R_1}\{p^{(1)}\}\chi_{R_2}\{p^{(2)}\}d_{R_3}\cdot\left<\chi_{_{R_1,R_2,R_3}}  \right>\stackrel{(\ref{Ga})}{=}\nn\\
=\sum_{R_1,R_2,R_3}\chi_{R_1}\{p^{(1)}\}\chi_{R_2}\{p^{(2)}\}C_{R_1,R_2,R_3}
{D_{R_1}(N_1)D_{R_2}(N_2)D_{R_3}(N_3)\over d_{R_1}d_{R_2}}
\ee
though such a partition function is anyway still not a generating function of all correlators with non-vanishing Gaussian averages, in accordance with (\ref{nr3}).
However, (\ref{nr3}) is already not reduced to a complex matrix model with some parameters $N_1$ and $N_2$ and, hence, is an interesting pattern to check integrability.

\paragraph{$W$-representation of tensor models.}

Note that one now construct a $W$-representation for the tensor model partition function (\ref{Z}). Indeed, let us note that if one constructs an operator $\hat O(N)$ with the property
\be\label{O}
\hat O(N) \chi_R = \frac{D_R(N)}{d_R} \chi_R
\ee
one can immediately obtain the partition function (\ref{Z}) acting with such operators on the generalized Cauchy formula (with using (\ref{Zr})):
\be\label{ZO}
Z_r\{p^{(i)}\}=\sum_{R_1,\ldots,R_r}C_{_{R_1\ldots R_r}} \prod_{m=1}^r\left(\chi_{R_m}\{p^{(m)}\}\cdot{D_{R_m}(N_m)
\over d_{R_m}}\right)=\nn\\
=\hat O_1(N_1)\ldots \hat O_m(N_m)\cdot
\sum_{R_1,\ldots,R_r}C_{_{R_1\ldots R_r}} \prod_{m=1}^r\chi_{R_m}\{p^{(m)}\}
=\hat O_1(N_1)\ldots \hat O_m(N_m)\cdot\exp\left(\sum_k
{\prod_{m=1}^rp^{(m)}_k\over k}\right)
\ee
where the subscript $m$ of $\hat O_m(N)$ means that this operator acts on the variables $p^{(m)}_k$.

In fact, such an operator $\hat O(N)$ has been constructed in \cite{AMMN2}, and is of the form
\be
\hat O(u)= N^{\hat W_{[1]}}\ {\sum_{\Delta}}^\prime N^{l(\Delta)-|\Delta|} \,\hat W_\Delta
\label{OvsW}
\ee
where sum goes over all diagrams containing no lines of unit length (we denote this restriction by prime)
and the differential operators (in $p_k$) $\hat W_\Delta$ are the standard generalized cut-and-join operators of \cite{MMN} ($\delta_i$ are the lengths of rows of the Young diagram $\Delta$):
\be
\hat W_{\!_\Delta}:= \frac{1}{z_{_\Delta}}:\prod_i \hat D_{\delta_i}:
\label{Wops}
\ee
and the derivative in the matrix $M$
\be
\hat D_k:= \Tr \left(M {\p\over \p M}\right)^k
\ee
acts on functions of the time variables $p_k = \Tr M^k$.
The normal ordering in (\ref{Wops}) implies that all the derivatives $\p_M$
stand to the right of all $M$.
Since $W_\Delta$ are gauge invariant matrix operators, and we apply them only to
gauge invariants, they can be realized as differential operators in $p_k$ \cite{MMN}.

It follows from (\ref{O}) that the operator $\hat O(u)$ preserves unity, $\hat O(u)\cdot 1 = 1$, which implies the relation
\be\boxed{
\begin{array}{c}
\displaystyle{
Z_r\{p^{(i)}\}=\hat O_1(N_1)\ldots \hat O_m(N_m)\circ\exp\left(\sum_k
{\prod_{m=1}^rp^{(m)}_k\over k}\right) \cdot 1=e^{\hat {\cal W}(N_1,...,N_r)}\cdot 1},\\
\\
\displaystyle{
\hat {\cal W}(N_1,...,N_r):= \hat O_1(N_1)\ldots \hat O_m(N_m)\circ
\sum_k{\prod_{m=1}^rp^{(m)}_k\over k}\circ \hat O_1^{-1}(N_1)\ldots \hat O_m^{-1}(N_m)}
\end{array}
}
\ee
using composition $\circ$ instead of action of operators, i.e. $\exp\left(\sum_k
{\prod_{m=1}^rp^{(m)}_k\over k}\right) $ is treated not as a function, but as an operator of multiplication. The differential operator (in $p^{(i)}_k$) $\hat {\cal W}(N_1,...,N_r)$ defining the $W$-representation of the tensor model
can be calculated similar to \cite{AMMN2}, where it was calculated in the matrix model case.

The operator $\hat O_1(N)$ is an element of $GL(\infty)$ \cite{AMMN2}, and, hence, it gives rise to a B\"acklund transformation: it maps a solution to the KP hierarchy to another solution.
In particular, acting on $\exp\left(\sum_k {\prod_{m=1}^rp^{(m)}_k\over k}\right)$, which is a trivial $\tau$-function still gives rise to a $\tau$-function w.r.t. $p^{(1)}_k$. Thus, we confirm that $\tau_0\{p^{(1)}_k,\ldots,p^{(r)}_k\}$ in (\ref{tau2}) is a $\tau$-function. However, it is no longer a $\tau$-function w.r.t. to other sets of time variables: applying $\hat O(N)$ destroys integrability w.r.t. to all other times. Hence, despite the product of $\hat O_i(N_i)$ at distinct $i$ belongs to the tensor product of a few $GL(\infty)$,
(\ref{ZO}) is no longer a $\tau$-function.

\paragraph{Conclusion.}
To conclude, we explained how superintegrability of the Gaussian
matrix models is lifted to the tensor level and discussed
integrability properties of associated Ooguri-Vafa partition functions.
Virasoro-like constraints are left beyond the scope of this letter,
because the {\it limited} integrability which we discussed
is already enough to demonstrate the need for a better choice
of partition function
and appeals for reexamining the longstanding problem of non-Abelian integrability
raised in \cite{GKLMM}.

\section*{Acknowledgements}

A. Mironov is grateful for the hospitality of NITEP, Osaka City University as well as that of the Workshop New Trends in Integrable Systems 2019 held there during the period of September, 9-20. Our work is partly supported by JSPS KAKENHI grant Number 19K03828 (H.I.), by the grant of the Foundation for the Advancement of Theoretical Physics ``BASIS" (A.Mir., A.Mor.), by  RFBR grants 19-01-00680 (A.Mir.) and 19-02-00815 (A.Mor.), by joint grants 19-51-53014-GFEN-a (A.Mir., A.Mor.), 19-51-50008-YaF-a (A.Mir.), 18-51-05015-Arm-a (A.Mir., A.Mor.), 18-51-45010-IND-a (A.Mir., A.Mor.). The work was also partly funded by RFBR and NSFB according to the research project 19-51-18006 (A.Mir., A.Mor.).


\begin{thebibliography}{12}

\bibitem{MM} A. Orlov, math-ph/0210012\\
A. Mironov, A. Morozov,
 Phys.Lett. {\bf B771} (2017)  503-507, arXiv:1705.00976; 	JHEP {\bf 2018} (2018) 163, arXiv:1807.02409

\bibitem{GMMMO} A. Gerasimov, A. Marshakov, A .Mironov, A. Morozov, A. Orlov,
Nucl.Phys. {\bf B357} (1991) 565

\bibitem{KMMOZ} S. Kharchev, A. Marshakov, A. Mironov, A .Orlov, A. Zabrodin,
Nucl.Phys. {\bf B366} (1991) 569-601

\bibitem{versus} S. Kharchev, A. Marshakov, A. Mironov, A. Morozov,
Nucl.Phys. {\bf B397} (1993) 339-378, hep-th/9203043

\bibitem{Mac} I.G. Macdonald, {\sl Symmetric functions and Hall polynomials},
Second Edition, Oxford
University Press, 1995

\bibitem{MPSh} A. Morozov, A. Popolitov, Sh. Shakirov, Phys.Lett. {\bf B784} (2018) 342-344, arXiv:1803.11401
\\
R. Lodin, A. Popolitov, Sh. Shakirov, M. Zabzine, arXiv:1810.00761

\bibitem{MMAGT} R. Dijkgraaf, C. Vafa, arXiv:0909.2453;\\
H. Itoyama, K. Maruyoshi, T. Oota,
Prog.Theor.Phys. {\bf 123} (2010) 957-987, arXiv:0911.4244\\
T. Eguchi, K. Maruyoshi,
arXiv:0911.4797;
arXiv:1006.0828\\
R. Schiappa, N. Wyllard,
arXiv:0911.5337\\
A. Mironov, A. Morozov, Sh. Shakirov,
JHEP {\bf 02} (2010) 030, arXiv:0911.5721;
Int.J.Mod.Phys. {\bf A25} (2010) 3173-3207, arXiv:1001.0563\\
H. Itoyama, T. Oota, Nucl. Phys. {\bf B838} (2010) 298-330, arXiv:1003.2929\\
A. Mironov, A. Morozov, An. Morozov, Nucl.Phys. {\bf B843} (2011) 534-557, arXiv:1003.5752

\bibitem{genM} A.~Morozov, A.~Smirnov,
   Lett.Math.Phys.\ {\bf 104} (2014) 585, arXiv:1307.2576\\
S.~Mironov, An.~Morozov, Y.~Zenkevich,
JETP Lett.\  {\bf 99} (2014) 109, arXiv:1312.5732 \\
Y.~Ohkubo, arXiv:1404.5401 \\
Y.~Kononov and A.~Morozov,  Eur.Phys.J. {\bf C76} (2016)  424,  arXiv:1607.00615     \\
Y.~Zenkevich,  arXiv:1612.09570

\bibitem{Nek} N. Nekrasov, Adv. Theor.Math.Phys. {\bf 7} (2004) 831-864, hep-th/0206161\\
R. Flume, R. Pogossian, Int.J.Mod.Phys. {\bf A18} (2003) 2541\\
N. Nekrasov, A. Okounkov, hep-th/0306238

\bibitem{CMM} A. Marshakov, A. Mironov, A. Morozov,
Phys. Lett. {\bf B265} (1991) 99\\
S. Kharchev, A. Marshakov, A. Mironov, A. Morozov, S. Pakuliak,
Nucl. Phys. {\bf B404} (1993) 717-750, hep-th/9208044\\
A. Mironov, S. Pakuliak, Theor.Math.Phys. {\bf 95} (1993) 604-625, hep-th/9209100\\
H.~Awata, Y.~Matsuo, S.~Odake, J.~Shiraishi,
  Phys.Lett. {\bf B347} (1995) 49, hep-th/9411053; Soryushiron Kenkyu {\bf 91} (1995) A69-A75, hep-th/9503028\\
H. Awata, Y. Matsuo, S. Odake, J. Shiraishi, Nucl.Phys. {\bf B449} (1995) 347-374, hep-th/9503043

\bibitem{MMSh} A. Mironov, A. Morozov, Sh. Shakirov, JHEP {\bf 1102} (2011) 067 arXiv:1012.3137\\
A. Mironov, A. Morozov, S. Shakirov, A. Smirnov, Nucl. Phys. {\bf B855} (2012) 128, arXiv:1105.0948

\bibitem{AGT} L.~Alday, D.~Gaiotto, Y.~Tachikawa,
  Lett.Math.Phys.\ {\bf 91} (2010) 167-197, arXiv:0906.3219\\
  N.~Wyllard,
  JHEP {\bf 0911} (2009) 002, arXiv:0907.2189\\
  A.~Mironov, A.~Morozov, Nucl.Phys.\ {\bf B825} (2009) 1-37,
  arXiv:0908.2569
  
  \bibitem{tenmods} R. Gurau; V. Rivasseau; S. Gielen, L. Sindoni; J.P. Ryan; V. Bonzom; S. Carrozza; T. Krajewski, R. Toriumi; A. Tanasa; SIGMA {\bf 12} (2016): ''Special Issue on Tensor Models, Formalism and Applications", http://www.emis.de/journals/SIGMA/Tensor\_Models.html\\
    Materials of the 2nd French-Russian Conference on Random Geometry and Physics
(2016), http://www.th.u-psud.fr/RGP16/

\bibitem{tenmodsSYK}
E. Witten, arXiv:1610.09758\\
R. Gurau, Nucl. Phys. {\bf B916} (2017) 386, arXiv:1611.04032; arXiv:1702.04228\\
I. Klebanov, G. Tarnopolsky, Phys.Rev. {\bf D 95} (2017) 046004, arXiv:1611.08915\\
H. Itoyama, A. Mironov, A. Morozov,
Phys.Lett. {\bf B771} (2017) 180-188, arXiv:1703.04983  \\
K. Bulycheva, I. Klebanov, A. Milekhin, G. Tarnopolsky,
arXiv:1707.09347\\
I.R. Klebanov, F. Popov, G. Tarnopolsky, PoS TASI2017 (2018) 004, arXiv:1808.09434\\
P. Diaz, J.A. Rosabal, JHEP {\bf 1901} (2019) 094, arXiv:1809.10153\\
F. Popov, arXiv:1907.02440\\
S. Prakash, R. Sinha, arXiv:1908.07178
  
\bibitem{Ramg1} R. de Mello Koch, S. Ramgoolam, arXiv:1002.1634 \\
D. Garner, S. Ramgoolam, Nucl.Phys. {\bf B875} (2013) 244-313, arXiv:1303.3246\\
J. Ben Geloun, S. Ramgoolam, arXiv:1307.6490;  arXiv:1806.01085\\
P. Diaz, S.J. Rey, arXiv:1706.02667,  arXiv:1801.10506\\
R. de Mello Koch, D. Gossman, L. Tribelhorn, JHEP {\bf 2017} (2017) 011, arXiv:1707.01455\\
P. Diaz,  arXiv:1803.04471\\
R.C. Avohou, J. Ben Geloun, N. Dub, arXiv:1907.04668

\bibitem{Ramg2} J. Ben Geloun, S. Ramgoolam, arXiv:1708.03524

\bibitem{IMMgcb} H. Itoyama, A. Mironov, A. Morozov, Phys.Lett. {\bf B788} (2019) 76-81, arXiv:1808.07783

\bibitem{IMMgc} H. Itoyama, A. Mironov, A. Morozov, arXiv:1909.06921 

\bibitem{IMMrain} H. Itoyama, A. Mironov, A. Morozov,
JHEP {\bf 2017} (2017) 1, arXiv:1704.08648; Nucl.Phys. {\bf B932} (2018) 52-118,  arXiv:1710.10027

\bibitem{Vir} F. David, Mod.Phys.Lett. {\bf A5} (1990) 1019\\
A. Mironov, A. Morozov, Phys.Lett. {\bf B252} (1990) 47-52\\
J. Ambj{\o}rn, Yu. Makeenko, Mod.Phys.Lett. {\bf A5} (1990) 1753\\
H. Itoyama, Y. Matsuo, Phys.Lett. {\bf 255B} (1991) 20

\bibitem{MMintrev} A. Morozov,
Phys.Usp.(UFN) {\bf 35} (1992) 671-714; {\bf 37} (1994) 1;
hep-th/9502091; hep-th/0502010\\
A. Mironov, Int.J.Mod.Phys. {\bf A9} (1994) 4355; Phys.Part.Nucl.
{\bf 33} (2002) 537; hep-th/9409190

\bibitem{GKLMM} A. Gerasimov, S. Khoroshkin, D. Lebedev, A. Mironov, A. Morozov,
Int.J.Mod.Phys. \textbf{A10} (1995) 2589-2614,
hep-th/9405011\\
A. Mironov, hep-th/9409190; Theor.Math.Phys. {\bf 114} (1998) 127, q-alg/9711006

\bibitem{FKN} M. Fukuma, H. Kawai, R. Nakayama, Int.J.Mod.Phys. {\bf A6} (1991) 1385

\bibitem{KS} V. Kac, A.S. Schwarz, 
Phys.Lett. {\bf B257} (1991) 329

\bibitem{GKM} S. Kharchev, A. Marshakov, A. Mironov, A. Morozov, A. Zabrodin,
Nucl.Phys. {\bf B380} (1992) 181-240, hep-th/9201013;
Phys.Lett. {\bf B275} (1992) 311-314, hep-th/9111037

\bibitem{MMZ2} A. Mironov, A. Morozov, Z. Zakirova, arXiv:1908.01278  

\bibitem{OV} H. Ooguri, C. Vafa, Nucl.Phys. {\bf B577} (2000) 419-438, arXiv:hep-th/9912123

\bibitem{SF} W. Fulton, {\sl Young tableaux: with applications to representation theory and geometry},
London Mathematical Society, 1997

\bibitem{MMten} A. Mironov, A. Morozov,
Phys.Lett. {\bf B774} (2017) 210-216, arXiv:1706.03667

\bibitem{DJKM} E. Date, M. Jimbo, M. Kashiwara, T. Miwa, 
RIMS Symp. {\sl "Non-linear integrable
systems - classical theory and quantum theory"} (World Scientific,
Singapore, 1983)

\bibitem{Sat} Y. Ohta, J. Satsuma, D. Takahashi, T. Tokihiro,
Prog.Theor.Phys.Suppl. {\bf 94} (1988) 210

\bibitem{gtau} S. Kharchev, A. Marshakov, A. Mironov, A. Morozov, Int.J.Mod.Phys. {\bf A10} (1995) 2015, hep-th/9312210\\
A. Orlov, D.M. Shcherbin,
Theor.Math.Phys.
{\bf 128} (2001) 906-926\\
A. Orlov,
Theor.Math.Phys. {\bf 146}
(2006) 183–206\\
A. Alexandrov, A. Mironov, A. Morozov, S. Natanzon,
J.Phys. A: Math.Theor. {\bf 45} (2012) 045209,
arXiv:1103.4100

\bibitem{AMMN2} A. Alexandrov, A. Mironov, A. Morozov, S. Natanzon,
JHEP {\bf 11} (2014) 080, arXiv:1405.1395

\bibitem{MMN} A. Mironov, A. Morozov, S. Natanzon,  Theor.Math.Phys. {\bf 166} (2011) 1-22,
arXiv:0904.4227;  Journal of Geometry and Physics {\bf 62} (2012) 148-155,
arXiv:1012.0433


\end{thebibliography}
\end{document}